\documentclass{article}
\usepackage[labelfont=bf]{caption} %
\usepackage[font={small}]{caption}   %

\usepackage{graphicx}%
\usepackage{dcolumn}%
\usepackage{bm}%
\usepackage[pdftex,colorlinks=true,linkcolor=blue,citecolor=blue]{hyperref}
\usepackage[utf8]{inputenc}
\usepackage{lipsum}  
\usepackage{amsmath}
\usepackage{authblk}
\usepackage{amsfonts}
\usepackage{placeins}
\usepackage[numbers, sort&compress]{natbib}  %
\usepackage{bibunits}
\usepackage{amssymb}
\usepackage{pdflscape} %
\usepackage{multirow}
\usepackage[percent]{overpic}
    \usepackage[table, dvipsnames]{xcolor}
    \usepackage{makecell}
    \usepackage{boldline}
    \setcellgapes{3pt}
\usepackage{amsmath}
\usepackage{amssymb}

\usepackage{textgreek}      %

\usepackage[strict]{changepage} %
\usepackage{nicefrac}

\defaultbibliographystyle{naturemag}
\defaultbibliography{references}

    \usepackage{soul}   %
    \setstcolor{red} %
    \setul{}{1.5pt}  %

\usepackage{lineno} 

\usepackage{paralist} 

\usepackage{xspace}

\title{Low loss monolithic barium titanate on insulator integrated photonics with intrinsic quality factor $>1$ million}

\author{
\large{
{Gwan In Kim$^{1, \dagger}$}, {Jieun Yim$^2$}, and {Gaurav Bahl$^{2}$}} \\
\vspace{12pt}
    \footnotesize{$^1$ Department of Electrical $\&$ Computer Engineering,} \\
    \footnotesize{$^2$ Department of Mechanical Science and Engineering,} \\
    \footnotesize{University of Illinois at Urbana–Champaign, Urbana, IL 61801 USA}

    \vspace{6pt}
    \footnotesize{$\dagger$ Corresponding author: {gikim2@illinois.edu}}
}

\date{}

\DeclareUnicodeCharacter{2212}{\textminus}
\begin{document}
\begin{bibunit}[naturemag]

\maketitle


\begin{abstract}

Barium titanate (BTO) has been experiencing a surge of interest for integrated photonics technologies because of its large nonlinear optical coefficients, especially the Pockels coefficient, and in part due to newly available thin-film substrates. In this work, we report on the development of a redeposition-free dry etching technique for monolithic BTO-on-insulator photonics, that produces very low-roughness and high-verticality waveguides. Using this, we experimentally demonstrate the first BTO microresonators with intrinsic Q-factor $> 1$ million, and waveguide propagation loss as small as 0.32 dB/cm, representing the lowest losses reported in any BTO-based integrated platform to date. We additionally demonstrate Mach-Zehnder amplitude modulators with $V_{\pi}L = 0.54$ V$\cdot$cm and effective electro-optic coefficient $r_\text{eff} = 162$ pm/V.

\end{abstract}

\vspace{12pt}

Barium titanate (BaTiO$_3$ or BTO) has drawn significant interest for electro-optic (EO) applications due to its exceptionally large Pockels coefficient (bulk measurement $r_{42} = 1640$~pm/V~\cite{johnston_determination_1965}) and there have been several efforts to bring this advantage to integrated photonics. 
The earliest attempts primarily focused on hybrid integration of BTO thin films with silicon \cite{abel_hybrid_2016, eltes_low-loss_2016, abel_large_2019, eltes_batio3-based_2019,posadas_rf-sputtered_2023,tao_towards_2024}, silicon nitride \cite{ortmann_ultra-low-power_2019, eltes_integrated_2020, posadas_thick_2021, kohli_plasmonic_2023} or silicon dioxide \cite{reynaud_microstructural_2022} to enable EO functionality without directly patterning the BTO layer. 
%
More recently, the emergence of BTO-on-insulator substrates~\cite{guo_epitaxial_2021, posadas_thick_2021} has opened the possibility of full optical confinement entirely within crystalline BTO waveguides. This shift toward monolithic platforms offers new opportunities to fully harness the Pockels coefficient and other non-linear optical properties of BTO \cite{petraru_ferroelectric_2002, dong_monolithic_2023,raju_highq_2025, chelladurai_barium_2025, riedhauser_absorption_2025, mohl_bidirectional_2025,isti_fabrication_2025}.

Presently, etching technology for BTO-on-insulator substrates for monolithic integrated photonics remains in an early stage. In particular, superior quality dry etches that produce smooth and highly vertical sidewalls are still underdeveloped, and yet are essential to minimize scattering and bending losses.
Initial attempts for BTO processing have relied on conventional fluorine or chlorine based dry etching chemistries \cite{li_comparative_2014, cheng_nanoscale_2020}. However, these processes tend to generate non-volatile barium fluoride and chloride byproducts, which adhere to the etched sidewalls resulting in surface roughness and increased optical scattering.
These byproducts are difficult to remove without high temperature processing \cite{cheng_nanoscale_2020}, and such methods are not conducive for BTO due to its low Curie temperature. 
An alternative approach is Ar$^+$ ion milling \cite{dong_monolithic_2023, raju_highq_2025}, which relies on physical ion bombardment, avoiding chemical byproducts and achieves good sidewall verticality.
However, here too it is common to find physical redeposition of sputtered material onto the etched sidewalls, which results in degraded surface quality. 
While similar redeposition occurs during Ar$^+$ etching of LiNbO$_3$ and LiTaO$_3$, it can be effectively removed through selective wet cleaning processes \cite{ulliac_argon_2016, wang_lithium_2024}. Unfortunately, an equivalent wet cleaning method for BTO has not yet been found. In the Supplementary \S\ref{sec:wetcleaning}, we present experimental examples of how common post-etch wet cleaning methods -- including RCA-1, hot piranha, KOH, and buffered oxide etch (BOE) -- were either found to cause BTO damage or were ineffective in removing the redeposited materials. Other recent studies have reported redeposition mitigation during etching \cite{mohl_bidirectional_2025, riedhauser_absorption_2025}, but this comes with the cost of reduced sidewall verticality, which exacerbates bending loss and compromises optical confinement.

\begin{figure}[ht!]
    \begin{adjustwidth*}{-1in}{-1in}
    \hsize=\linewidth
    \includegraphics[width=0.9\columnwidth]{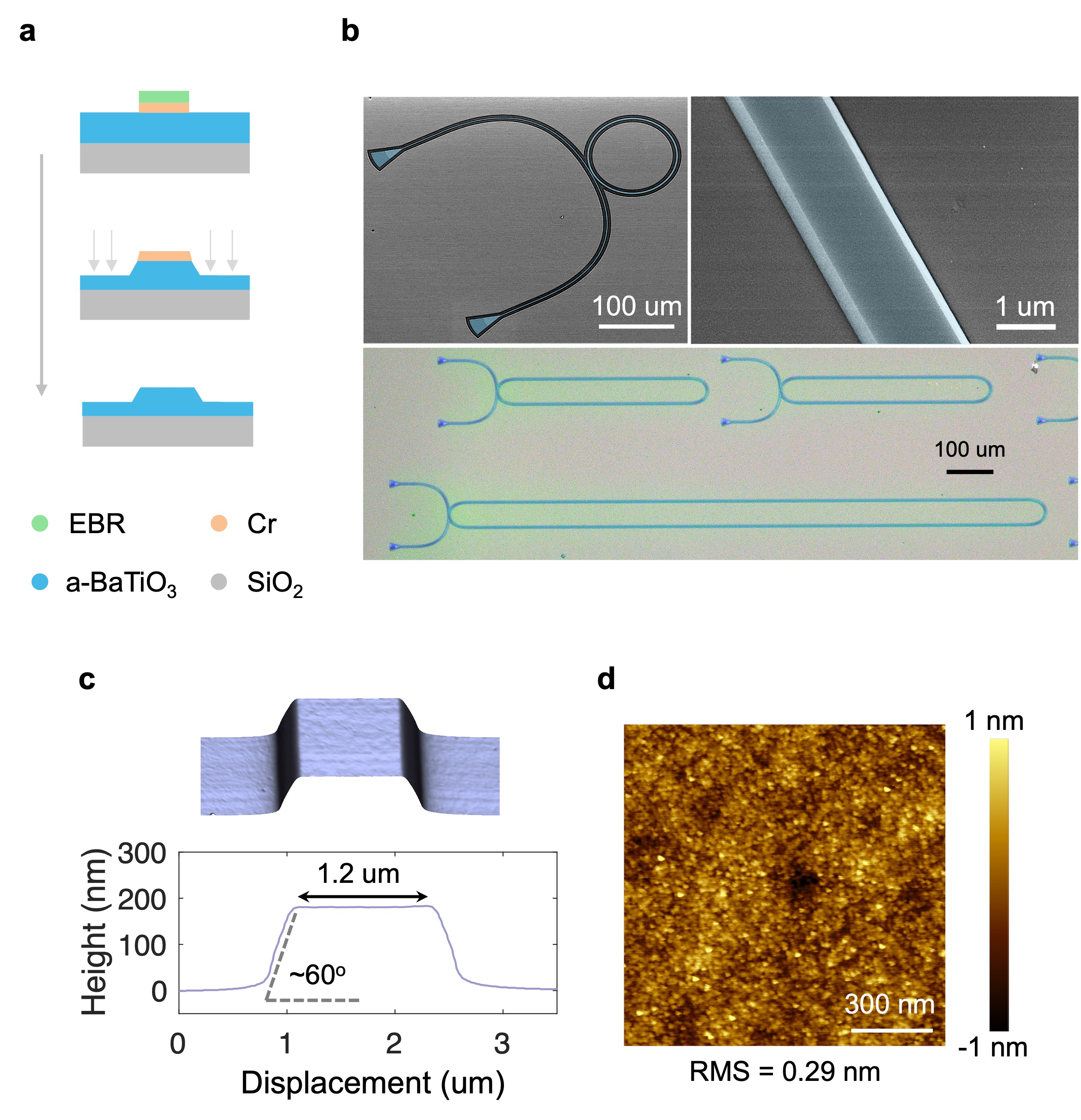}
    \centering
    \caption{
        \textbf{Low loss thin-film barium titanate on insulator integrated photonics.}
        \textbf{(a)}
        Schematic fabrication flow for monolithic BTO ridge waveguide devices. Electro-beam lithography on ZEP-520A e-beam resist (EBR) is used to define a Cr hardmask, followed by BTO etching via Ar$^+$ ion milling (with small fraction of Cl$_2$ gas), and then Cr removal using a wet etch.
        \textbf{(b)}
        Scanning electron microscope (SEM) image of monolithic BTO microring resonator and waveguide. Bottom optical microscope image shows BTO microracetrack resonator.
        \textbf{(c)}
        Atomic force microscope (AFM) scan image of our BTO-on-insulator waveguide, with a 1D slice showing a sidewall angle nearly $60^\circ$.
        \textbf{(d)}
        2D AFM scan of the etched surface shows etched surface roughness of $\approx 0.29$ nm rms.
    }
    \label{fig1}
    \end{adjustwidth*}
\end{figure}

In this work, we report the development of a fabrication process for monolithic BTO-on-insulator photonic integrated circuits, without any undesirable residues, very low surface roughness, and near-optimal sidewall geometry. Using this, we experimentally demonstrate the highest resonator intrinsic quality factor ($Q_i >$ 1 million) and lowest propagation loss (0.32 dB/cm) ever measured in any BTO integrated photonics platform.

\vspace{12pt}

Our optimized etching process uses Ar$^+$ ion milling as the primary physical etch for BTO, and a small fraction of Cl$_2$ gas is introduced to chemically remove the sputtered material into volatile byproducts which are pumped out of the chamber without forming residue. This process improvement is inspired by an approach that was explored for Ar$^+$ etching of lithium niobate {\cite{younesi_fabrication_2025}}. The starting substrate is 340 nm thick a-oriented BTO film on 3 {\textmu}m thick SiO\textsubscript{2} on a silicon handle wafer (Fig.~\ref{fig1}a). A chromium (Cr) hard mask was deposited and patterned using 150 keV electron beam lithography (EBL, Elionix ELS-G150) with ZEP-520A resist. The exposed BTO was then etched to a depth of 180 nm at an etch rate of $\approx 17.5$ nm/min using inductively coupled plasma reactive ion etching (ICP-RIE), followed by Cr mask removal.

An ideal waveguide etch profile must exhibit smooth sidewalls and high verticality to ensure optical confinement. To validate the optimized etch process, we present scanning electron microscope (SEM) images of the microring resonator and zoom-in view of waveguide region (Fig.~\ref{fig1}b), which show smooth, redeposition-free sidewalls. Further assessment using atomic force microscopy (AFM) and height profile analysis (Fig.~\ref{fig1}c,d) reveals a trapezoidal waveguide cross-section with a top width of $\approx 1.2$ {\textmu}m, sidewall angles of $\approx 60^\circ$, and an etch depth of 180 nm. A 2D AFM scan indicates an rms surface roughness of 0.29 nm, confirming excellent surface smoothness conducive to low optical scattering and bending loss.

\begin{figure}[htp]
\begin{adjustwidth*}{-1in}{-1in}
  \centering
  \begin{overpic}[width=1.0\textwidth,tics=10]{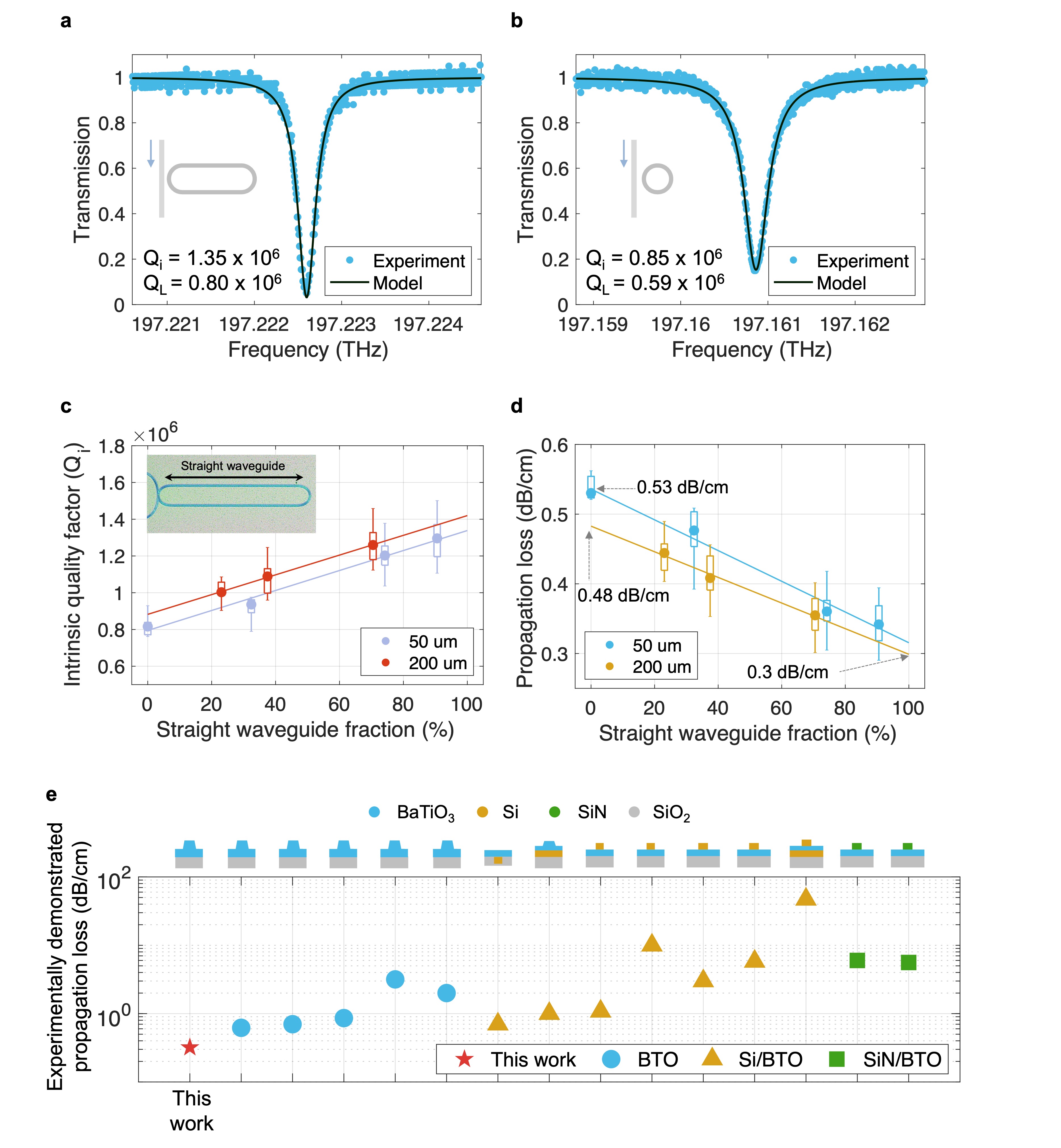}
    \put (18.7,3.0) {\small$\displaystyle\cite{raju_highq_2025}^\dag$}
    \put (23.7,3.0) {\small$\displaystyle\cite{isti_fabrication_2025}$}
    \put (27.9,3.0) {\small$\displaystyle\cite{riedhauser_absorption_2025}$}
    \put (32.4,3.0) {\small$\displaystyle\cite{dong_monolithic_2023}$}
    \put (37.0,3.0) {\small$\displaystyle\cite{petraru_ferroelectric_2002}$}
    \put (42.0,3.0) {\small$\displaystyle\cite{tao_towards_2024}$}
    \put (46.7,3.0) {\small$\displaystyle\cite{posadas_rf-sputtered_2023}$}
    \put (51,3.0) {\small$\displaystyle\cite{abel_large_2019}$}
    \put (55.1,3.0) {\small$\displaystyle\cite{eltes_batio3-based_2019}_1$}
    \put (59.5,3.0) {\small$\displaystyle\cite{eltes_batio3-based_2019}_2$}
    \put (64.7,3.0) {\small$\displaystyle\cite{abel_hybrid_2016}$}
    \put (68.9,3.0) {\small$\displaystyle\cite{eltes_low-loss_2016}$}
    \put (73.5,3.0) {\small$\displaystyle\cite{eltes_integrated_2020}$}
    \put (78.0,3.0) {\small$\displaystyle\cite{ortmann_ultra-low-power_2019}$}
  \end{overpic}
   \caption{
        \textbf{Optical characterization of monolithic BTO-on-insulator microresonators.}
        \textbf{(a)}
        Resonance line width and quality factor analysis of microracetrack resonator with 200 $\mu$m radius with 70.5\% straight waveguide fraction, measured near 1520 nm. 
        \textbf{(b)}
        Quality factor analysis of microring resonator with 50 $\mu$m radius. 
        \textbf{(c)}
        Statistical analysis of measured intrinsic quality factor ($Q_i$) for resonators with varying straight waveguide fraction, in the range 1520--1537.4 nm (195--197.3 THz). See text for further details. For each set of measurements, the dot shows the mean value, the box represents the interquartile range, and the whiskers denote the minimum and maximum values. 
        \textbf{(d)}
        Statistical analysis and extrapolation of the propagation loss based on the measurements in \textbf{(c)}, leading to an estimated 0.3 dB/cm straight waveguide loss.
        \textbf{(e)} Benchmarking of the measured propagation loss values 
        from this work and previously reported BTO-based integrated photonics, including monolithic BTO-on-insulator \cite{isti_fabrication_2025,raju_highq_2025, riedhauser_absorption_2025, dong_monolithic_2023, petraru_ferroelectric_2002}, Si/BTO hybrid~\cite{tao_towards_2024, posadas_rf-sputtered_2023, abel_large_2019,eltes_batio3-based_2019, abel_hybrid_2016, eltes_low-loss_2016}, and SiN/BTO~\cite{eltes_integrated_2020, ortmann_ultra-low-power_2019} hybrid. Details are provided in Supplementary Table~\mbox{\ref{FOM_table}}. $\dag$ Estimated from the data shown in the publication.       
    }
    \label{fig2}
\end{adjustwidth*}
\end{figure}
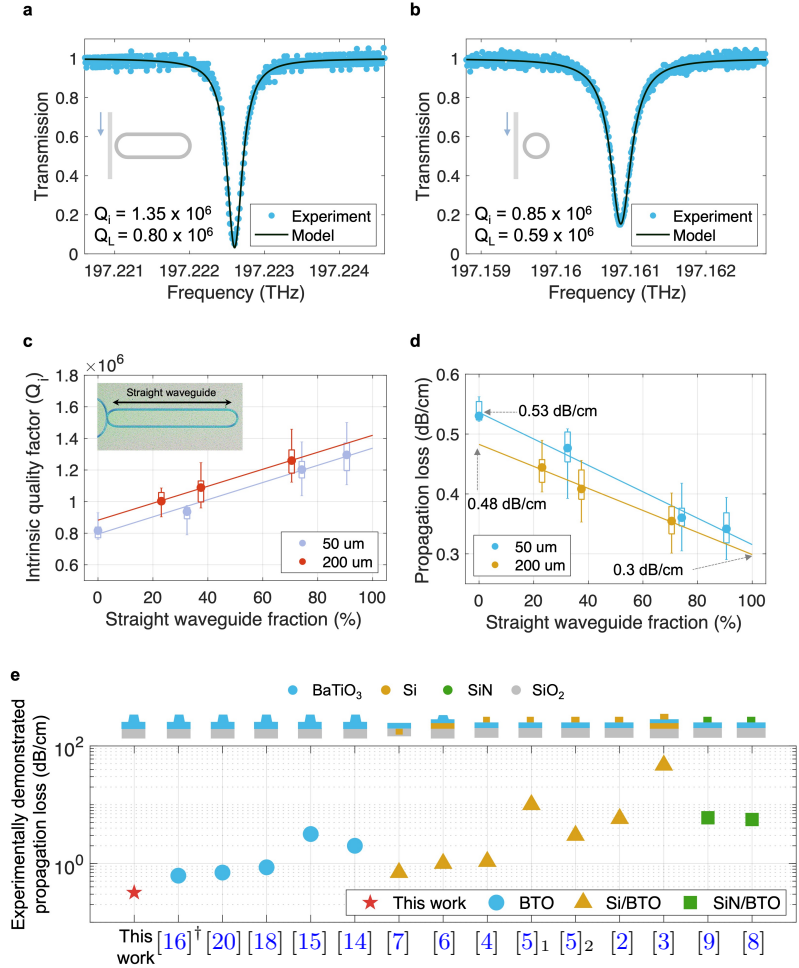

Fig.~\ref{fig2} presents a comprehensive optical characterization of the fabricated microracetrack and ring resonators, based on measurements taken in the telecom band.
First, Fig.~\ref{fig2}a shows a representative transmission spectrum through the bus waveguide of a 200~$\mu$m-turn radius microracetrack resonator, measured near 1520 nm, displaying the resonance dip of near-critical coupling. For this device we measure intrinsic quality factor $Q_i = 1.35 \times 10^6$ and loaded quality factor $Q_L = 0.8 \times 10^6$. 
To the best of our knowledge, this is the first demonstration of an optical microresonator with $Q_i$ exceeding 1 million, and the highest quality factors (both intrinsic and loaded) in any BTO-based integrated photonics platform. This includes monolithic BTO-on-insulator \mbox{\cite{riedhauser_absorption_2025, mohl_bidirectional_2025, raju_highq_2025}} as well as hybrid structure that incorporate Si~\mbox{\cite{eltes_batio3-based_2019,abel_hybrid_2016}}. In {Fig.~\ref{fig2}b} we present a representative transmission spectrum (also measured near 1520 nm) for a microring resonator with a tighter bending radius of 50 $\mu$m.  For this device we estimate quality factors of $Q_i = 0.84 \times 10^6$ and $Q_L = 0.59 \times 10^6$.

In order to determine the linear waveguiding loss, independent of curvature-induced losses, we fabricated a series of microracetrack resonators with varying straight waveguide fraction and two options for turn radius (50~$\mu$m and 200~$\mu$m).
Fig.~{\ref{fig2}}c presents the experimentally extracted intrinsic quality factor $Q_i$ as a function of the straight waveguide fraction. Fig.~{\ref{fig2}}d re-maps this information to an equivalent propagation loss via the relation $\alpha = 2\pi n_g/(Q_i \lambda) \cdot 10 \log_{10} e$ (dB/cm), where $n_g$ is group index and $\lambda$ is optical wavelength.
Measurements for all devices were performed over the range 1520--1537.4 nm (195–197.3~THz), with between 7 to 35 resonances contributing to each data point depending on the device free spectral range. The error bars in Fig.~{\ref{fig2}}c,d represent the the statistical variation over all resonances measured for that device.

We find that as the straight waveguide fraction increases, $Q_i$ also increases, confirming that propagation loss is lower in straight sections compared to curved sections. Notably, the 200~$\mu$m racetracks consistently exhibit higher $Q_i$ values across all devices, which can be attributed to the larger bend radius and reduced curvature-induced loss. 
Through linear fit in Fig.~{\ref{fig2}}d we are able to separately extrapolate the loss contributions from straight and curved segments. We therefore estimate the straight waveguide loss to be $\alpha \approx 0.3$~dB/cm, 
while the bending loss is extracted as $\approx 0.53$~dB/cm for the 50~$\mu$m bend radius and $\approx 0.48$~dB/cm for the 200~$\mu$m bend radius. 
Our best individual result within the statistical data is $\alpha = 0.32~\mathrm{dB/cm}$, corresponding to the $Q_i = 1.35 \times 10^6$ individual resonator measurement of Fig.~\mbox{\ref{fig2}a}.

To benchmark these results, Fig.~\ref{fig2}e presents a comparison of measured propagation losses reported from various BTO photonic platforms, including monolithic BTO-on-insulator \cite{isti_fabrication_2025,raju_highq_2025,riedhauser_absorption_2025, dong_monolithic_2023, petraru_ferroelectric_2002} and hybrid integrations with Si~\cite{tao_towards_2024, posadas_rf-sputtered_2023, abel_large_2019, eltes_batio3-based_2019, abel_hybrid_2016, eltes_low-loss_2016} and SiN\textsubscript{x}~\cite{kohli_plasmonic_2023, eltes_integrated_2020, ortmann_ultra-low-power_2019} (details are provided in Supplementary Table~\mbox{\ref{FOM_table}}).
Our best result $\alpha = 0.32$~dB/cm represents the lowest measured propagation loss to date for any BTO-based integrated photonics platform, i.e. without relying on extrapolations.
We attribute this improvement to the significantly reduced optical loss enabled by smooth and highly vertical etched sidewalls, which control both scattering and bending losses.

\vspace{12pt}

\begin{figure}[t!]
    \begin{adjustwidth*}{-1in}{-1in}
    \hsize=\linewidth
    \includegraphics[width=1.0\columnwidth]{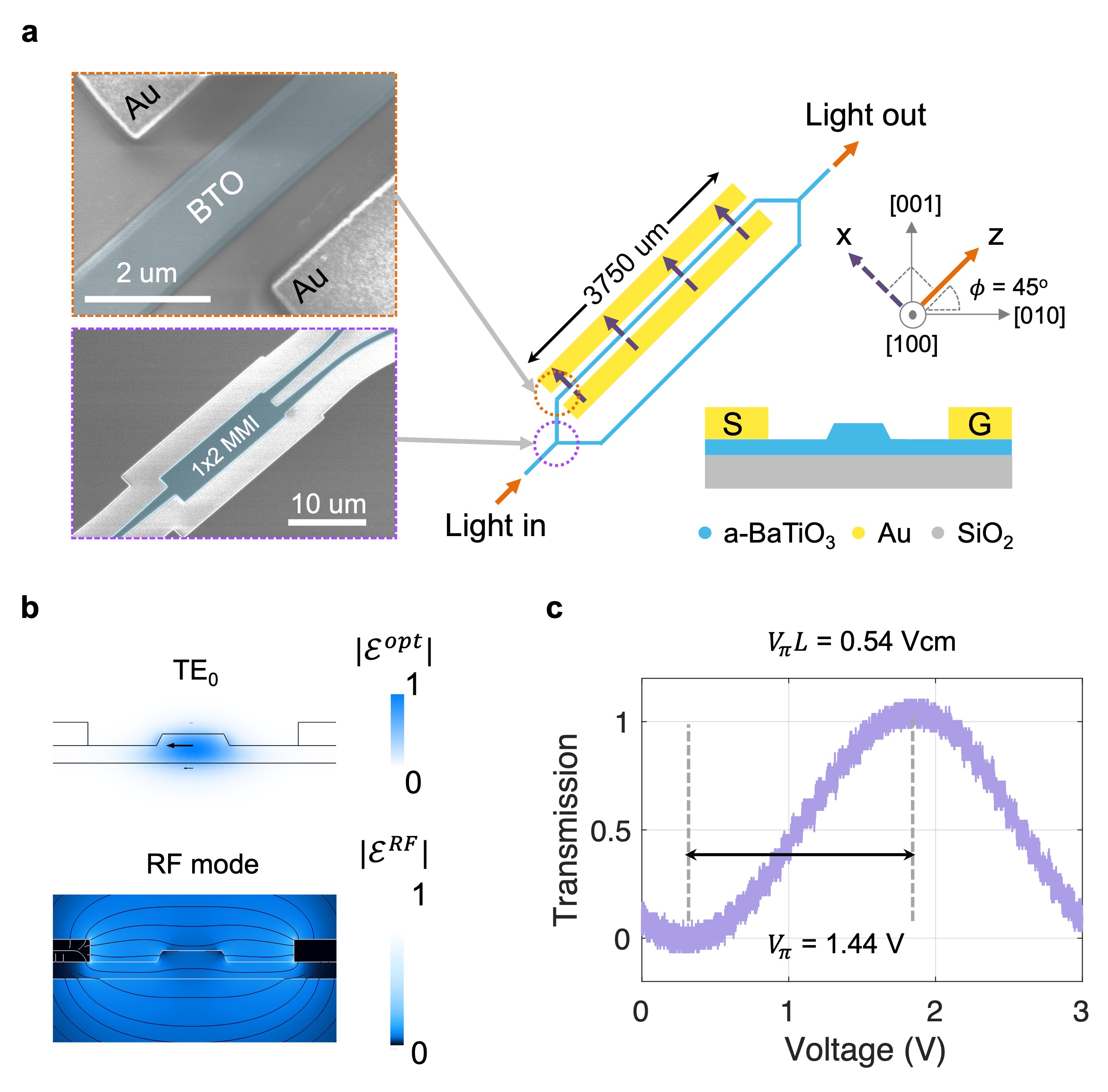}
    \centering
    \caption{
        \textbf{Electro-optic modulation analysis}
        \textbf{(a)}
        Schematic diagram of the Mach-Zehnder modulator (MZM) that was fabricated in thin-film BTO-on-insulator substrate.
        \textbf{(b)}
        Finite element analysis simulation of optical TE\textsubscript{0} mode and RF mode within MZM.
        \textbf{(c)}
       Top-view SEM image of MZM. Bottom left shows 1x2 multi mode interferometer (MMI) and bottom right image shows EO modulator with waveguide and Au electrode.
        \textbf{(d)}
        Measured optical transmission as a function of voltage, indicating $V_{\pi} = 1.44$ V and $V_{\pi}L = 0.54$ V$\cdot$cm.
        }
    \label{fig3}
    \end{adjustwidth*}
\end{figure}

A major motivation for monolithic BTO photonics is the potential for highly efficient electro-optic (EO) modulators and optical switches. To explore this in our platform, we fabricated simple Mach–Zehnder modulators (MZMs) as shown in Fig.~{\ref{fig3}}. 
We focused our design on optical TE mode waveguides in the a-orientation BTO-on-insulator substrate (i.e., with the [100] axis crystal orientation out-of-plane), with simple electrodes that produce a transverse electric field as shown in Fig.~{\ref{fig3}a}. 
Since the Pockels tensor is highly anisotropic, this combination produces an effective coefficient {\cite{chelladurai_barium_2025}} expressed as $r_{\text{eff}}(\phi) = r_{33} \cos^3 \phi + (r_{13} + 2r_{42}) \sin^2 \phi \cos \phi$, where $\phi$ is the angle chosen between the optical propagation axis and the crystallographic [010] axis, as illustrated in Fig.~{\ref{fig3}}a. The effective $r_{\text{eff}}$ is maximized near $\phi \approx 54^\circ$.
However, since the BTO ferroelectric domains are typically split into two orthogonal predominant in-plane orientations {\cite{demkov_ferroelectric_2024}}, the best practical effect is obtained at $\phi = 45^\circ$ with lateral RF electric field along the \mbox{$[0\overline{1}1]$} direction.

We designed the single-arm MZM interferometer (Fig.~\ref{fig3}a) with arm length 3.75 mm oriented along the $[011]$ axis of the BTO crystal. The gap between the stimulus electrodes was set to 3.6 $\mu$m.
We pole the BTO by applying dc voltage of 120~V across the stimulus electrodes for 30 minutes prior to measurement, corresponding to a maximum E-field of 333 kV/cm (Fig.~\ref{fig3}b). In Fig.~{\ref{fig3}}c we present the measured transmission through the MZM as a function of applied voltage across the electrodes, showing the expected sinusoidal dependence due to phase modulation within the interferometer. 
From this data, we extract a half-wave voltage of $V_\pi = 1.44$~V and $V_\pi \times L = 0.54$ V$\cdot$cm.

We can now estimate the effective Pockels coefficient through the equation
	$ r_\text{eff}= \lambda\, g / (n^3\, \Gamma \, V_{\pi}L) $
where 
$V_{\pi}L$ is obtained from the MZM experimental measurements, $n$ is refractive index of BTO ($\approx 2.26$ at 1550 nm), $g$ is the electrode gap, and $\Gamma$ is the overlap integral between optical mode and electric field \cite{dong_monolithic_2023, chelladurai_barium_2025}.
Here we calculate $r_\text{eff} \approx 162$ pm/V, which is significantly higher than the largest Pockels tensor element available in lithium niobate, and comparable to past works on monolithic BTO-on-insulator photonics~\cite{dong_monolithic_2023, mohl_bidirectional_2025, chelladurai_barium_2025}. Since the electro-optic response of BTO is sensitive to the polarization state and domain orientation, which requires proper poling \cite{posadas_effect_2024}. Therefore, the effective $r_\text{eff}$ could further improved by implementing enhanced poling techniques, such as with dedicated sidewall electrodes \cite{franken_milliwatt-level_2025}. Such improvements could further unlock the full potential of the Pockels effect in BTO for integrated photonics.

\vspace{12pt}

The development of high quality material growth, patterning, and etching processes has been key to the enormous success of integrated photonics platforms using Si, SiN$_x$, LiNbO$_3$, LiTaO$_3$, and will similarly be the key to the successful development of monolithic BTO-on-insulator technologies. 
The progress we have demonstrated is a major enabling step for a wide range of high-performance photonic applications requiring large nonlinear optical coefficients and long photon lifetimes, including frequency combs (electro-optic and Kerr combs)~\mbox{\cite{zhang_broadband_2019, delhaye_optical_2007}}, higher harmonic generation~\mbox{\cite{moore_continuous-wave_2011}}, narrow-linewidth lasers~\mbox{\cite{jin_hertz-linewidth_2021}}, and squeezed light sources~\mbox{\cite{safavi-naeini_squeezed_2013}}. 
The programmability advantage of BTO, owing to its characteristic as a soft ferroelectric, can be additionally valuable for optical computing~\cite{ahmed_universal_2025, ou_hypermultiplexed_2025} and reconfigurable photonics~\cite{liu_fully_2016, prabhathan_roadmap_2023}.

\nolinenumbers   

\vspace{24pt}

\FloatBarrier

\end{bibunit}
%
\section*{Acknowledgments}

The authors would like to acknowledge invaluable support from Dr.~Edmond Chow and Dr.~Soo Ho Choi at the Holonyak Micro \& Nanotechnology Lab at the University of Illinois at Urbana-Champaign, from Dr.~Agham Posadas at La Luce Cristallina, and from Dr.~Brian Kasch at the Air Force Research Laboratory (AFRL) Space Vehicles Directorate.
This work was sponsored by the AFRL / US Space Force grant FA9453-20-2-0001, the US Office of Naval Research (ONR) Multi-University Research Initiative grant N00014-20-1-2325, and the Defense Advanced Research Projects Agency (DARPA) under Cooperative Agreement D24AC00003. 
The views and conclusions contained herein are those of the authors and should not be interpreted as necessarily representing the official policies or endorsements, either expressed or implied, of DARPA, AFRL, ONR, or the US Government.

\section*{Data availability}

The data that support the findings of this study are available from the corresponding authors upon reasonable request.

\section*{Author contributions}

G.I.K. and G.B. jointly conceived this work. G.I.K. performed the theoretical analysis, performed the device fabrication, and conducted the experiments. G.I.K. and J.Y. analyzed the experimental data. All authors contributed to writing the manuscript. G.B. supervised all aspects of this project.

\FloatBarrier

\newpage

\renewcommand*{\citenumfont}[1]{S#1}
\renewcommand*{\bibnumfmt}[1]{[S#1]}
\newcommand{\beginsupplement}{%
        \setcounter{table}{0}
        \renewcommand{\thetable}{S\arabic{table}}%
        \setcounter{figure}{0}
        \renewcommand{\thefigure}{S\arabic{figure}}%
        \setcounter{equation}{0}
        \renewcommand{\theequation}{S\arabic{equation}}%
        \setcounter{section}{0}
        \renewcommand{\thesection}{S\arabic{section}}%
}

\beginsupplement

\begin{bibunit}
\begin{center}

\Large{\textbf{Supplementary Information: \\ Low loss monolithic barium titanate on insulator integrated photonics with intrinsic quality factor $>$ 1 million}} \\
\vspace{12pt}
\vspace{12pt}
\large{
{Gwan In Kim$^1$}, {Jieun Yim$^2$}, and {Gaurav Bahl$^2$}} \\
\vspace{12pt}
    \footnotesize{$^1$ Department of Electrical $\&$ Computer Engineering,} \\
    \footnotesize{$^2$ Department of Mechanical Science and Engineering,} \\
    \footnotesize{University of Illinois at Urbana–Champaign, Urbana, IL 61801 USA} \\
\end{center}

\section{Evaluation of wet cleaning strategies for redeposition removal}
\label{sec:wetcleaning}

On-sidewall redeposition of etch products is a known issue in Ar$^+$ ion milling, as it relies on physical ion bombardment. This undesirable redeposition along the etched sidewalls introduces significant roughness and degrades optical propagation loss. While such redeposition is often addressed by post-etch wet cleaning steps in other materials, such as LiNbO$_3$ and LiTaO$_3$, there is not yet a comparable method available for BTO.

\begin{figure}[htp]
    \begin{adjustwidth*}{-1in}{-1in}
    \hsize=\linewidth
    \includegraphics[width=0.9\columnwidth]{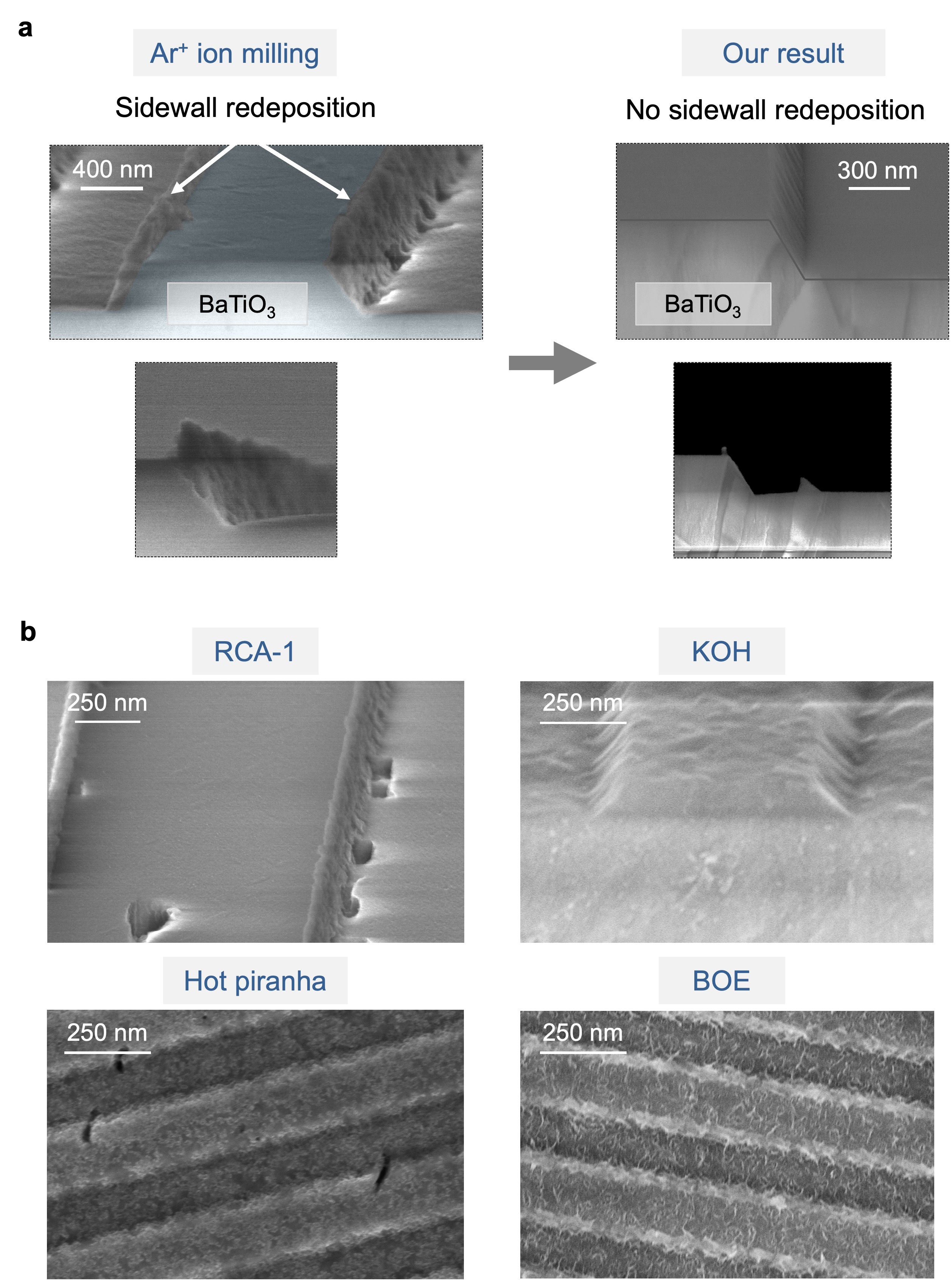}
    \centering
    \caption{
            \textbf{Comparison of sidewall redeposition and wet cleaning effects on BTO waveguides.} 
            (a) Cross-sectional SEM images comparing typical Ar$^+$ ion milling (left) with our optimized etching process (right). Conventional ion milling results in significant sidewall redeposition, while our process yields smooth, residue-free sidewalls. 
            (b) SEM images showing the effect of various wet cleaning treatments on BTO sidewalls after redeposition. RCA-1 (top left) is not effective and introduces roughness. KOH:H$_2$O$_2$ (top right) causes topographical damage. Hot piranha (bottom left) leads to severe surface roughening, and BOE (bottom right) fails to eliminate redeposition and causes additional morphological distortion.
    }
    \label{wetcleaning}
    \end{adjustwidth*}
\end{figure}

To make a comparison, we show a BTO etching result using our optimized dry etching process in Fig.~\ref{wetcleaning}a on the right. This approach yields smooth BTO sidewalls with no observable etch product redeposition. We can compare this to a more conventional Ar$^+$ ion milling process performed on BTO (Fig.~\ref{wetcleaning}a on the left), whose the etch recipe is similar to the lithium niobate version \cite{orsel_electro-optic_2023, sohn_electrically_2021}, and we observe prominent physical redeposition along the sidewalls. 

To evaluate whether various wet cleaning techniques can remove the redeposited material, we tested several commonly used wet cleaning chemistries -- including RCA-1, hot piranha, KOH:H$_2$O$_2$, and buffered oxide etchant (BOE). As shown in Fig.~\ref{wetcleaning}b, RCA-1 cleaning (NH$_4$OH:H$_2$O$_2$:H$_2$O = 1:1:5, 40 minutes) still leaves substantial redeposition along the sidewalls and introduces surface pinholes. Hot piranha treatment causes pronounced surface etching and morphological damage to the BTO, likely due to its aggressive oxidizing nature. KOH cleaning, which is effective for LiTaO$_3$~\cite{wang_lithium_2024}, leads to surface roughening, which is also similarly shown in BOE (6:1) treatment.

These observations collectively suggest that wet cleaning chemistries optimized for other ferroelectric oxides (LiNbO\textsubscript{3} or LiTaO\textsubscript{3}) do not translate effectively to BTO. In many cases, they exacerbate surface roughness or induce damage, which is detrimental to low-loss photonic device performance. Therefore, achieving redeposition-free etch profiles in BTO is a welcome strategy.

\clearpage

\vspace{24pt}

\clearpage
\FloatBarrier

\vspace{60pt}

\newcommand{\NA}[0]{\cellcolor{black!15}N/A}
\newcommand{\TFBT}[0]{\cellcolor{RoyalBlue!15}BTOI}
\newcommand{\SiNBT}[0]{\cellcolor{pink!15}SiN/BTO}
\newcommand{\SiBT}[0]{\cellcolor{yellow!15}Si/BTO}
\newcommand{\SiOBT}[0]{\cellcolor{green!15}SiO\textsubscript{2}/BTO}

\begin{landscape}
		\begin{table}[h]
		\small

			\vspace{-60pt}
			\caption{
                Comparison of measured 
                quality factors and propagation losses (i.e. without extrapolation) in various BTO integrated photonics platforms.
                }
			
			\begin{adjustwidth}{0.3in}{-0.5in}
			\begin{tabular}{| >{\raggedright\arraybackslash}m{3.3cm} | >{\raggedright\arraybackslash}m{0.9cm} | >{\raggedright\arraybackslash}m{2cm} | >{\centering\arraybackslash}m{1.6cm} | >{\raggedright\arraybackslash}m{2.5cm} | >{\raggedright\arraybackslash}m{2.5cm} | >{\raggedright\arraybackslash}m{2.5cm} | >{\raggedright\arraybackslash}m{1.8cm} | >{\raggedright\arraybackslash}m{1.8cm} | >{\raggedright\arraybackslash}m{1.8cm} | >{\raggedright\arraybackslash}m{1.6cm} |}
				
				\hline
				Publication & Year & Device Type & Platform & Intrinsic Q-factor ($Q_{i}$) & Loaded Q-factor ($Q_{L}$) & Propagation loss ($\alpha$,~dB/cm) & Wavelength (nm)   \\ \hline \hline

                This work &  & Racetrack Resonator & \TFBT & 1.35 x 10\textsuperscript{6} & 0.80 x 10\textsuperscript{6} & 0.32 & 1520\\
                 &  & Ring Resonator & \TFBT & 0.85 x 10\textsuperscript{6} & 0.59 x 10\textsuperscript{6} & 0.52 & 1520\\ \hline \hline

                Isti et. al.~\cite{isti_fabrication_2025}& 2025 & Waveguide & \TFBT & \NA & \NA & 0.7 & 1546  \\ \hline
                Raju et. al.~\cite{raju_highq_2025}& 2025 & Racetrack Resonator & \TFBT & 0.5 x 10\textsuperscript{6} & 0.09 x 10\textsuperscript{6} & 0.62$^\dag$ & 1550 \\
                & & Ring Resonator & \TFBT & 0.2 x 10\textsuperscript{6} & 0.10 x 10\textsuperscript{6} & 2.11 & 1550\\
                \hline
                Riedhauser et. al.~\cite{riedhauser_absorption_2025}& 2025 & Resonator & \TFBT & 0.45 x 10\textsuperscript{6} & \NA & 0.86 & 1554  \\ \hline
                M\"ohl et. al.~\cite{mohl_bidirectional_2025}& 2025 & Resonator & \TFBT & 0.25 x 10\textsuperscript{6}$^\dag$ ($\kappa_i$ = 776 MHz at 1550 nm) & \NA & \NA & 1537--1577  \\ \hline
                Dong et. al.~\cite{dong_monolithic_2023}& 2023 & Waveguide & \TFBT & \NA & \NA & 3.17 & 1535--1565\\ \hline
                Petraru et. al.~\cite{petraru_ferroelectric_2002}& 2002 & Waveguide & \TFBT & \NA & \NA & 2--3 & 632  \\ \hline
                \hline

                Tao et. al.~\cite{tao_towards_2024}& 2023 & Waveguide & \SiBT & \NA & \NA & 1 & 1545  \\ \hline
                Posadas et. al.~\cite{posadas_rf-sputtered_2023}& 2023 & Waveguide & \SiBT & \NA & \NA & 1.08 & 1550  \\ \hline
                Abel et. al.~\cite{abel_large_2019}& 2019 & Waveguide & \SiBT & \NA & \NA & 10 & 1541 \\ \hline
                Eltes et. al.~\cite{eltes_batio3-based_2019}& 2019 & Waveguide & \SiBT & \NA & \NA & 3 & 1550--1551  \\ 
                & & Resonator & \SiBT & \NA & 0.05 x 10$^6$ & 5.8 &   \\ \hline 
                Abel et. al.~\cite{abel_hybrid_2016}& 2016 & Resonator & \SiBT & \NA & 0.0046 x 10$^6$ & 47 & 1555--1575   \\ \hline
                Eltes et. al.~\cite{eltes_low-loss_2016}& 2016 & Waveguide & \SiBT & \NA & \NA & 6 & 1550  \\ 
                \hline
                \hline
                Kohli et. al.~\cite{kohli_plasmonic_2023}& 2023 & Waveguide (plasmonic) & \SiNBT & \NA & \NA & 5000 & 1557 \\ \hline
                Eltes et. al.~\cite{eltes_integrated_2020}& 2020 & Waveguide & \SiNBT & \NA & \NA & 5.6 & 1511--1512\\ \hline
                Ortmann et. al.~\cite{ortmann_ultra-low-power_2019}& 2019 & Resonator & \SiNBT & \NA & \NA & 9.4 & 1550 \\ \hline

			\end{tabular}
			\end{adjustwidth}
			
			\vspace{12pt}

			\label{FOM_table}
			
			BTOI = Thin-film barium titanate (BTO)-on-insulator.  Si/BTO = Silicon and barium titanate hybrid.  SiN/BTO = Silicon nitride and barium titanate hybrid. N/A = Information not available or not applicable.\\
            $\dag$ Estimated from the data shown in the publication.
            
		\end{table}
    
\end{landscape}
\clearpage
\FloatBarrier

\renewcommand{\refname}{Supplementary References}

\end{bibunit}

\end{document}